\documentclass[twocolumn,showpacs,preprintnumbers,amsmath,amssymb]{revtex4}

\usepackage{graphicx}
\usepackage{dcolumn}
\usepackage{bm}
\usepackage[raggedright,nooneline]{subfigure}

\begin{document}

\preprint{APS/123-QED}

\title{Asymmetric Two-dimensional Magnetic Lattices for Ultracold Atoms Trapping and Confinement}

\author{A. Abdelrahman}
 \email{a.abdelrahman@ecu.edu.au} 
 \altaffiliation{%
\\Electron Science Research Institute, Edith Cowan University,  270 Joondalup Drive, Joondalup WA 6027 Australia}
\author{ M. Vasiliev}%
\affiliation{%
Electron Science Research Institute, Edith Cowan University,  270 Joondalup Drive, Joondalup WA 6027 Australia}%
\author{K. Alameh}%
\affiliation{%
Electron Science Research Institute, Edith Cowan University,  270 Joondalup Drive, Joondalup WA 6027 Australia}%
\author{P. Hannaford}
\email{Phannaford@swin.edu.au}
\affiliation{Centre  for Atom Optics and Ultrafast Spectroscopy, and ARC Centre of Excellence for Quantum Atom Optics,
Swinburne University of Technology, Melbourne, Australia 3122}%

\date{\today}

\begin{abstract}
 A new method to implement an asymmetrical two-dimensional magnetic lattice is proposed. The asymmetrical two-dimensional magnetic lattice can be created by periodically distributing magnetic minima across the surface of magnetic thin film where the periodicity can be achieved by milling $n\times n$ square holes on the surface of the film. The quantum device is proposed for trapping and confining ultracold atoms and quantum degenerate gases prepared in the low magnetic field seeking-state at low temperature, such as the Bose-Einstein Condensate (BEC) and ultracold fermions. We present detailed analysis of the analytical expressions and the numerical simulation procedure used to calculate the external magnetic field. We also, describe the magnetic band gap structure exhibited by the asymmetric effect of the magnetic minima and show some of the possible  application. We analyze the effect of changing the characteristic parameters of the magnetic lattice, such as the separating periodicity length and the hole size along with the applications of the external magnetic bias fields to maintain and allocate a suitable non-zero magnetic local minima at effective $z$-distance above the thin film surface. Suitable values are shown which keep the trapped ultracold atoms away from the thermal Majorana spin-flip and the surface Casimir-Polder effect.    
\end{abstract} 
\pacs{Valid PACS appear here}
\maketitle

\section{\label{sec:1}INTRODUCTION}

The emerging field of quantum computing and information processing using ultracold atoms prepared in specific quantum states and trapped in specifically patterned magnetic minima, has attracted a great interest over the last decade where magnetic micro-traps have been recently recognized to have more applicable accessibility to several degrees of freedom of their individual hosted quantum systems \cite{Ref_15}. Several micrometer-scale structures generating magnetic periodic potential have been proposed for such purpose as an alternative to optical lattices \cite{Ref_1}. It is possible to produce a lattice configuration by periodically introducing specific patterns on the surface of a magnetic material, where in such configuration they are known as magnetic lattices \cite{Ref_2,Ref_3,Ref_4,Ref_5,Ref_9}, in which they can be realized by manufacturing one-dimensional and two-dimensional periodically distributed magnetic minima. Such specifically engineered quantum devices are also classified as Atom Chips \cite{Ref_6,Ref_7,Ref_8}. Magnetic lattices can  also be created using current carrying wire, however a few technical difficulties have been encountered in relation to the integration of current-carrying wires onto the magnetic micro-traps. For example, current-produced near-field thermal noise reduces the number of trapped cold atoms. Also, the current density fluctuations lead to the fragmentation in the BEC cloud \cite{Ref_10,Ref_11}. These technical difficulties may be avoided by using optimized magnetic materials possessing high remanent magnetization and enabling highly stable trap frequencies through a tight magnetic confinement and reproducible periodic potentials \cite{Ref_12}. It may also offer a controllable phase transitions, possible manipulation and less demanding access to the quantum ensemble for measurement \cite{Ref_18}. Also, the ability of magnetic lattices to attain pronounced confinement and a proper control of single particles makes them excellent candidates to host large-scale qubit systems \cite{Ref_1,Ref_13,Ref_17,Ref_19}. In this article, we propose, a simple method to implement an asymmetric two-dimensional magnetic lattice, using the current micro-fabrication technologies, where we analyze the performance of the magnetic lattice for ultracold atoms confinement and trapping, with the aid of applying in situ external magnetic bias fields that are integrated on the atom chip. In section (\ref{sec:2}) we derive the analytical expressions which describe the magnetic minima distribution across the surface of the magnetic thin film. In section (\ref{sec:3}) we show the simulation procedure used to calculate the surface effective magnetic field and in section (\ref{sec:4}) we describe the relevant characteristic parameters that will be used to optimize the magnetic minima distribution and hence preparing suitable quantum states such as two-level systems. We conclude in section (\ref{sec:5}) by showing the possible application of this type of magnetic lattice.

\section{\label{sec:2}Magnetic Lattices Structure}

\begin{figure*}
\numberwithin{figure}{section}
\subfigure[][]{
\label{Fig1a}}
\subfigure[][]{
\label{Fig1b}}
\subfigure[]{
\label{Fig1c}}
\subfigure[]{
\label{Fig1d}}
\subfigure[]{
\label{Fig1e}}
\subfigure[]{
\label{Fig1f}}
\subfigure[]{
\label{Fig1g}}
\subfigure[]{
\label{Fig1h}}
\caption{\label{Fig1} (a) A $10 \times 10$ magnetic lattice surrounded by an unperturbed area. (b) The lattice parameters are specified by the hole size $\alpha_h \times \alpha_h$, the periodic spacing $\alpha_s$ between the holes and the magnetic layer thickness $\tau_{btm}$. (c) Magnetic density plot of the simulated magnetic lattice sites at $z-x$ plane along the center of the lattice. The traps (dark color) are located at the effective $z$-distance, $d_{min}$, from the holes opening centers of high magnetic field (bright color) above the film plane. (d-e) Magnetic field density plot across the $x-y$ plane at the $d_{min}$ with no external magnetic bias fields and with application of the external $B_{x-bias}$ = 10 G and $B_{y-bias}$ = 10 G magnetic bias fields, respectively. (f-g) Contour plot of distributed lattice sites across the $x-y$ plane without and with applications of $B_{x-bias}$ = 10 G and $B_{y-bias}$ = 10 G respectively. (h) 3D plot of the magnetic field (on the z-axis) of the distributed sites across the $x-y$ plane ($x$-axis and $y$-axis) simulated at the $d_{min}$ and displayed from the center sites to the edge sites where $B_{x-bias}$ = 10 G and $B_{y-bias}$ = 10 G were applied. Simulation input: $\alpha_s=\alpha_h$ = 1 $\mu m$ $M_z$ = 3 kG, $\tau_{btm}$ = 2 $\mu m$ and $\tau_{p-wall}$  = 1 $\mu m$.}
\end{figure*}

The proposed structure to generate the magnetic lattice based quantum device is realized by milling an $m\times m$ array of blocks, each block being an array of $n\times n$ square holes of size $\alpha_h$, separated by $\alpha_s$, where $n$ represents the number of holes in a block as detailed in Figures \ref{Fig1}(a-b). The gaps between the blocks containing no holes are assumed to be greater than, or equal in width to $\alpha_s$. The holes are formed, within a magneto-optic thin film of thickness $\tau_{btm}$ sputtered onto a proper substrate. The depths of all holes are equal and extend through the thin film thickness down to the substrate surface level. The gaps separating different blocks are an important design feature which enables the $n\times n$ magnetic lattices to be surrounded by an unperturbed film area, which introduces an extra degree of confinement through the creation of magnetic field "walls" encircling the $n\times n$ matrices, and isolating them from one another as shown in Figure \ref{Fig1}(h). The magnetic quantum device is in its remanently-magnetized state, where its magnetization, $M_z$, is directed in a direction perpendicular to the surface plane. Effective parameters, i.e., $\alpha_h$, $\alpha_s$ and $\tau_{btm}$ of the magnetic lattice for different values of $n$ are considered.

The structure generates two dimensional periodically distributed magnetic field minima in the vicinity of the surface of the perforated film where the distribution creates the magnetic lattice that used to trap the cold atoms. The presence of holes results in a magnetic field distribution whose non-zero local minima are located at effective z-distances from the holes opening centers above the film plane, denoted here by $d_{min}$. These minima are localized in small volumes representing the potential wells that contain certain number of quantized energy levels for the cold atoms to occupy. In our design, we  assumed that the size of the holes $\alpha_h$ and the holes separation $\alpha_s$ are both equal, $\alpha_h$ = $\alpha_s$ $\equiv$ $\alpha$, to simplify the mathematical derivation where we adopted an analysis approach similar to that reported in \cite{Ref_3,Ref_4}. The spatial magnetic field components $B_x$, $B_y$ and $B_z$ can be written analytically as a combination of a field decaying away
from the surface of the trap in the $z$-direction and a periodically distributed magnetic field in the $x-y$ plane produced only by  the magnetic induction, $B_o = \mu_oM_z/\pi$, at the surface of the permanently magnetized thin film. We define the surface reference magnetic field as $B_{ref} = B_o(1-e^{-\beta \tau})$, where $\beta = \pi/\alpha$, $\tau = \tau_{btm}$ is the film thickness, and the plane of symmetry is assumed at $z = 0$. Although our model takes into account the magnetic local minima created only by $B_o$ (similar to a flat magnetic mirror), we also include in the analysis the external magnetic bias field components along $x, y$ and $z$ directions, namely, $B_{x-bias}$, $B_{y-bias}$ and $B_{z-bias}$ which can also be localized by means of micro-fabrications. Figure \ref{Fig1}(d-e)-\ref{Fig1}(f-g) show the simulated map of the magnetic field strength distribution across the $x-y$ planes located at different distances above the magnetized film surface. Both the density plot and contour plot representation types are shown for both simulation results of a $10\times 10$ magnetic lattice at the initial magnetic state, formed by $B_{ref}$ only, and with the external application of $B_{x-bias}$ and $B_{y-bias}$, respectively.
The analytical expressions that describe the non-zero local minima, periodically positioned in the $x-y$ plan, take into account the strength of the effective field and $\alpha$  \cite{Ref_4}. These expression are derived and simplified to the following set of equations 
\begin{widetext}
\begin{equation}
B_{x} = B_{o}\biggr(1-e^{-\beta\tau}\biggl)e^{-\beta[z-\tau]}sin(\beta x)-\frac{B_{o}}{3}\biggr(1-e^{-3\beta\tau}\biggl)\times e^{-3\beta[z-\tau]}sin(3\beta x)+\ldots+B_{x-bias}\label{eq:1}
\end{equation}
\begin{equation}
B_{y} = B_{o}\biggr(1-e^{-\beta\tau}\biggl)e^{-\beta[z-\tau]}sin(\beta y)-\frac{B_{o}}{3}\biggr(1-e^{-3\beta\tau}\biggl)e^{-3\beta[z-\tau]} \times sin(3\beta y)+\ldots+B_{y-bias} \label{eq:2}
\end{equation}
\begin{equation}
B_{z} = B_{o}\biggr(1-e^{-\beta\tau}\biggl)e^{-\beta[z-\tau]}\biggl[cos(\beta x)+cos(\beta y)\biggl]-\frac{B_{o}}{3}\biggr(1-e^{-3\beta\tau}\biggl)e^{-3\beta[z-\tau]}\biggl[cos(3\beta x)+cos(3\beta y)\biggl]+\ldots+B_{z-bias}\label{eq:3-1}
\end{equation}
\end{widetext}
The higher order terms in these equations, which describe the far field, can be neglected for distances that are large compared to the effective distance. This is because
only the cold atoms that are in the so called low magnetic field seeking-state (atom's magnetic moment to be oriented antiparallel
to the localized magnetic field in the trap) are attracted to the non-zero local minima of the magnetic traps located at the effective
$z$-distance from the film surface, which is larger than $\alpha/2\pi$ . Thus, Equations (\ref{eq:1}) to (\ref{eq:3-1}) are reduced to the followings
\begin{widetext}
\begin{eqnarray}
B_{x} & = & B_{o}\biggr(1-e^{-\beta\tau}\biggl)e^{-\beta[z-\tau]} sin(\beta x) + B_{x-bias}\label{eq:4}\\
B_{y} & = & B_{o}\biggr(1-e^{-\beta\tau}\biggl)e^{-\beta[z-\tau]} sin(\beta y)+ B_{y-bias}\label{eq:5}\\
B_{z} & = & B_{o}\biggr(1-e^{-\beta\tau}\biggl)e^{-\beta[z-\tau]} \biggl[cos(\beta x)+cos(\beta y)\biggr]+ B_{z-bias}\label{eq:6}
\end{eqnarray}
\end{widetext}
The magnitude $\textbf{B}$ of the magnetic field above the film surface can be written, using Equations (\ref{eq:4}) to (\ref{eq:6}), as

\begin{multline}
\textbf{B} =  \Biggl\{ B_{x-bias}^{2}+B_{y-bias}^{2}+B_{z-bias}^{2}\\
              + 2B_{o}^{2}\biggl(1-e^{-\beta\tau}\biggr)^{2}e^{-2\beta[z-\tau]}\biggl[cos(\beta x)cos(\beta y)\biggr]\\
              + 2B_{o}^{2}\biggl(1-e^{-\beta\tau}\biggr)e^{-\beta[z-\tau]}\Biggl(\biggl[B_{x-bias}+ B_{z-bias}\biggl]cos(\beta x)\\
              + \biggl[B_{y-bias}+B_{z-bias}\biggl]cos(\beta y)\Biggl)\Biggl\}^{1/2} \label{eq:7}
\end{multline}
\begin{figure*}
\numberwithin{figure}{section}
\subfigure[]{
\label{Fig2a}}
\subfigure[]{
\label{Fig2b}}
\subfigure[]{
\label{Fig2c}}
\subfigure[]{
\label{Fig2d}}
\caption{\label{Fig2}(a) The magnetic field minima $B^z_{min}$ along the $z$-axis at the effective $z$-distance, $d_{min}$, away from the surface of the thin film. (b) The field minima calculated along $z$-axis at the center and the edge lattice sites. (c) Magnetic field distribution simulated along the x-axis for a $20\times 20$ two-dimensional magnetic lattice.(d) Two magnetic quantum wells at the lattice edge along the $y$-axis separated by the magnetic tunneling barrier $\Delta B$ and have different values of magnetic non-zero local minima through tilted (magnetic) potential $\Delta V$.  Simulation input of the magnetic lattice characteristic parameters: $M_z$ = 3 kG and $\tau_{btm}$ = 2 $\mu m$ with no external bias fields applied. }
\end{figure*}
The simulation result of numerical calculations for a finite magnetic lattice are shown in Figure \ref{Fig2}. Figure \ref{Fig2}(a) shows the location of the magnetic field non-zero local minima along the $z$-axis, $B^z_{min}$, at the effective distance ,$d_{min}$, from the hole opening center and confined along $z$-axis by a magnetic barrier of magnitude  $\Delta B_z$. The results shown in Figure \ref{Fig2}(b) demonstrate the existence of non-zero local minima of the magnetic field at the effective $z$-distance along the $x$-axis, confirming the suitability of the structure proposed for trapping the atoms being in the low magnetic field seeking state. The intra-block magnetic lattices consist of tunneling barriers separating the trapping sites where each magnetic quantum well characterized by the non-zero local minima and differs from each other by amount of tilted potential $\Delta V$ as shown in Figure \ref{Fig2}(d).  The unperturbed film area, around the $n\times n$ hole-matrix block, produces magnetic walls that localize and magnetically decouple the individual blocks from each other, as illustrated in Figure \ref{Fig2}(c). This allows individual quantum-state addressing for each $n\times n$ magnetic lattice separately, where different quantum features can be investigated, such as quantum multi-particle (multipartite) long-range entanglement  and strong correlations \cite{Ref_19}, the large scale lattice spacing in our magnetic trap is suitable for single atoms detection \cite{Ref_14}. It is also possible to vary the individual tunneling barrier heights by applying an external magnetic field in the negative $z$-direction. The numerical calculations in Figure \ref{Fig3}(c-d) show the effect of applying small external $B_{z-bias}$ magnetic fields on the tunneling barrier across the lattice sites. The z-bias field is possible to be applied uniformly from outside the vacuum system of the BEC experiment. It is also possible in case of large $\alpha_s$ to  control the tunneling barriers in situ through micro-coils surrounding each hole.
Table (\ref{Table1}) summarize the important parameters used in simulating the proposed magnetic lattice-based quantum device, where $\tau_{btm}$,
$\tau_{wall}$ represent the bottom layer and the wall thickness of the two dimensional magnetic lattice, respectively. Table (\ref{Table2}) shows
the output of the two-dimensional magnetic lattice characteristic parameters describing the suitability of using the quantum device to provide a longer lifetime for the trapped ultracold atoms to process quantum information.

\begin{table*}
\caption{Simulation parameters used to produce the non-zero local minima $B_{min}$with
their effective distances $d_{min}$. The magnetic walls are
 produced by the condition $\tau_{wall}\geqslant\tau_{btm}$.}
\begin{ruledtabular}
\begin{tabular}{cccccccc}
Simulation parameters & $m\times m$ & $n\times n$ & $M_{z}$ & $\tau_{btm}$ & $\tau_{wall}$ & $\alpha_{h}$ & $\alpha_{s}$\tabularnewline
\hline 
Input values & $1\times1$ & $11\times 11$ & 2 kG &  2 $\mu m$ & 1 $\mu m$ & 1 $\mu m$ & 1 $\mu m$\tabularnewline
\end{tabular}
\end{ruledtabular}
\label{Table1}
\end{table*}
\begin{table*}
\caption{Characteristic parameters describe the suitability of the magnetic
lattice to host the cold atoms. The magnetic walls are produced by
the condition $\tau_{wall}\geqslant\tau_{btm}$. Edge site is denoted by $B_{min}^{A}$, while $B_{min}^{B}$ denotes center site.}
\begin{ruledtabular}
\begin{tabular}{cccccc}
Magnetic lattice parameters & $B_{x-bias}$ & $B_{min}^{A}$ & $B_{min}^{B}$ & $d_{min}^{A}$ & $d_{min}^{B}$\tabularnewline
\hline 
Simulation output & 0 G & 0.37 G & 0.5 G & 0.8451 $\mu m$ & 0.718 $\mu m$\tabularnewline

 & 2 G & 2.03 G & 2.1 G & 0.8451 $\mu m$ & 0.717 $\mu m$\tabularnewline
\end{tabular}
\end{ruledtabular}
\label{Table2}
\end{table*}
\begin{figure*}
\numberwithin{figure}{section}
\subfigure[][]{
\label{Fig3a}}
\subfigure[][]{
\label{Fig3b}}
\subfigure[][]{
\label{Fig3c}}
\subfigure[][]{
\label{Fig3d}}
\caption{\label{Fig3} (a) The effect of changing the size of the holes, $\alpha_h$, on the location of the magnetic field local minima along the $z$-axis, $B^z_{min}$ located at $d_{min}$ above the hole opening center at the surface of the thin film, and (b) the effect of changing the periodicity length, $\alpha_s$, across the $x-y$ plane. (c) Simulation result of variating the tunneling barrier heights $\Delta B_x$ through change in the $z$-axis magnetization, $M_z$ by applying external  $B_{z-bias}$ magnetic field in the negative z-direction, and (d) $B_{z-bias}$ effect on the gradient of the magnetic sites near the local minima along the $z$-axis . The thin film thickness: $\tau_{btm}$ = 2 $\mu m$ is used in all cases.}
\end{figure*}

\section{\label{sec:3} Simulation of the surface magnetic field of the magnetic lattices}
 To simulate the magnetic lattices, field-producing volumes are represented by symmetrical segments, where each segment is subdivided into a suitable number of smaller objects for accurate computation. Uniform magnetization $M_{z}$ is assumed to exist within each segment of the total volume of the $n\times n$ magnetic lattices, which produces the magnetic field distribution in the space surrounding the structure. Mutual interactions between the segmented volumes are accounted for, and all relevant data are stored in a relatively large matrix. The simulation is carried out through the relaxation procedure by first loading the material properties, the magnetization $M_{z}$, the layer thickness $\tau=\tau_{btm}$, and the structure geometry which is done by specifying the values of $\alpha_{h}$, $\alpha_{s}$, and $n$. The external surface  magnetic field is then calculated at different points in the space around the surface of the magnetized thin film. The computed fields at the non-zero local minima then provide accurate estimates for the depths of the potential wells and their locations at the effective $z$-distance. The numerical results are calculated using the package RADIA \cite{Ref_16}.

We simulate the important effects of changing the structure dimensions on the magnetic lattice characteristic parameters, in particular the location of the non-zero local minima, $d_{min}$, from the surface of the thin film, their sizes and the curvature of the magnetic field across each individual magnetic potential well that forms a lattice site. The gradient of the magnetic field can be controlled via change in $\alpha_h$, also shallow or steeper magnetic potential wells can be realized when choosing the suitable dimensions of $\alpha_s$ and $\alpha_h$. Figure (\ref{Fig3}) summarize the simulation results of changing the hole size $\alpha_h$, the length of the periodicity $\alpha_s$ and applications of the external magnetic bias field along the negative direction of the $z$-axis, $B_{-z-bias}$, and their effects on the relevant characteristic parameters including the height of the tunneling magnetic barrier $\Delta B$ and the amount of the potential tilt $\Delta V$ between two adjacent sites of the magnetic lattice.

\begin{figure*}
\numberwithin{figure}{section}
\subfigure[][]{
\label{Fig4a}}
\numberwithin{figure}{section}
\subfigure[][]{
\label{Fig4b}}
\subfigure[][]{
\label{Fig4c}}
\subfigure[][]{
\label{Fig4d}}
\caption{\label{Fig4}(a-d) Size of the magnetic potential wells along the $x$-axis for different sizes of holes. The simulation is carried out using the initial conditions $M_z$ = 2 kG, $\tau_{btm}$ = 2 $\mu m$ and with no external bias fields applied  for the different values  $\alpha_h=\alpha_s$ = 1 $\mu m$, 3 $\mu m$, 5 $\mu m$ and 7 $\mu m$.}
\end{figure*}

\section{\label{sec:4} Characteristic Parameters of the Asymmetrical Two-Dimensional Magnetic Lattice}
The initialization setup of the magnetic lattice, using only the magnetic induction at the surface of the permanently magnetized thin film in the $z$ direction, leads to a symmetrical distribution of the potential wells in the $x-y$ plane. We simulate this effect by applying the constrains $B_{x-bias}=B_{y-bias}=B_{z-bias}=0$ and $B_{o}\approx \mu_{o}M_{z}/\pi$, where the dominant factor is the surface reference magnetic field
$B_{ref}$ . Applying the above constrains, Equation (\ref{eq:7}) reduces to 
\begin{equation}
\textbf{B}=B_{ref}e^{-\beta[z-\tau]}\sqrt{2cos(\beta x)cos(\beta y)}\label{eq:8}
\end{equation}

The asymmetric distribution of the non-zero local minima is periodically spaced, and the magnetic field minima, $B_{min}$, located at points defined by the coordinates ($x_{min},y_{min},d_{min}$), where $d_{min}$ determine the effective $z$-distances of $x-y$ plane that contains the distributed lattice sites. Mainly the parameters $B_{min}$ and $d_{min}$, determine the life time of the trapped cold atoms which make it critical to locate the $d_{min}$ and choose the suitable value of $B_{min}$ when micro-fabricating the quantum device. The minima locations, along the $x$, $y$ and $z$-axis, respectively, can be written as 

\begin{eqnarray}
x_{min} & = & n_{x}\alpha,\qquad\qquad\qquad n_{x}=0,\pm1,\pm2,\cdots\label{eq:9}\\
y_{min} & = & n_{y}\alpha(n_{x}),\qquad\qquad n_{y}=0,\pm2,\pm4,\cdots\label{eq:10}\\
d_{min} & \approx & \frac{\alpha}{\pi}ln\Biggl(B_{ref}+\frac{1}{B_{z-bias}}\Biggr)\label{eq:11}
\end{eqnarray}
\begin{figure*}
\numberwithin{figure}{section}
\subfigure[]{
\label{Fig5a}}
\subfigure[]{
\label{Fig5b}}
\subfigure[]{
\label{Fig5c}}
\numberwithin{figure}{section}
\subfigure[]{
\label{Fig5d}}
\subfigure[]{
\label{Fig5e}}
\subfigure[]{
\label{Fig5f}}
\subfigure[]{
\label{Fig5g}}
\subfigure[]{
\label{Fig5h}}
\caption{\label{Fig5}(a) Schematic diagrams of the conditions positive wall, $\tau_{p-wall}> \tau_{btm}$. (b) Density plot representation of the simulation results of applying the condition $\tau_{p-wall}$ = 0.5 $\mu m$ ($\tau_{p-wall}> \tau_{btm}$) on the 9 $\times$ 9 magnetic lattice which causes reductions in the tilting magnetic potential between the lattice sites. (c) Comparing the magnetic minima locations between the center site and the edge site along the $z$-axis. (d) A $100 \times 100$ magnetic lattice simulation using the $\tau_{p-wall}$ = 1 $\mu m$ constrain. Random distribution of the sites appears at the edges of the magnetic lattice. (e) Schematic representation of the condition negative walls, $\tau_{n-wall}< \tau_{btm}$, where (f) the density plot representation shows the effect of the condition $\tau_{n-wall}$ = -0.5 $\mu m$ on the locations of the sites along the $z$-axis which are spaced by the tilting potential. (h) For large magnetic lattice the effect disappears from the center sites while robustly appears at the edges. As simulation parameters we used $M_z$ = 2.8 kG, $\tau_{btm}$ = 3 $\mu m$ and no external bias fields has been applied. $\alpha_h = \alpha_s$ = 7 $\mu m$ and $\alpha_h = \alpha_s$ = 10 $\mu m$ is used for large lattice size, i.e., $n=100$.}
\end{figure*}
\begin{table*}
\label{Table3}
\caption{Simulation results show the effect of changing the periodicity $\alpha_s$ of a $11\times 11$ asymmetric two-dimensional magnetic lattice prepared initially at the reference magnetic field $B_{ref}$ and no external magnetic bias fields are applied.}
\begin{ruledtabular}
\begin{tabular}{cccccccc}
$M_{z}(kG)$ & $\alpha_{h}(\mu m)$ & $\alpha_{s}(\mu m)$ & $B_{min}^{center}(G)$ & $d_{min}^{center}(\mu m)$ & $B_{min}^{edge}(G)$ & $d_{min}^{edge}(\mu m)$ & $\tau_{wall}$ $condition$\tabularnewline
\hline
\hline 
$3.8$ & $0.5$ & $0.5$ & $0.3$ & $0.2516$ & $0.91$ & $0.2909$ & $\tau_{se-wall}$\tabularnewline
\hline 
$2.0$ & $0.5$ & $1.5$ & $1.5273$ & $0.448$ & $3.218$ & $0.4878$ & $\tau_{se-wall}$\tabularnewline
\hline 
$2.0$ & $1$ & $1$ & $0.5$ & $0.669$ & $0.6$ & $0.7467$ & $\tau_{se-wall}$\tabularnewline
\hline 
$2.0$ & $1$ & $2$ & $2.15$ & $0.811$ & $2.802$ & $0.947$ & $\tau_{p-wall}$\tabularnewline
\end{tabular}
\end{ruledtabular}
\end{table*}
\begin{figure*}
\numberwithin{figure}{section}
\subfigure[]{
\label{Fig6a}}
\subfigure[]{
\label{Fig6b}}
\subfigure[]{
\label{Fig6c}}
\subfigure[]{
\label{Fig6d}}
\caption{\label{Fig6}The effect of applying the $B_{x-bias}$ field on the magnetic tunneling barrier $\Delta B_x$, $\Delta B_z$ and the non-zero magnetic local minima $B^x_{min}$ (a) along  $x$-axis and (b) along $z$-axis for different values of $B_{x-bias}$. (c) Results of applying external bias field $B_{+z-bias}$ along $z$-axis, and (d) the effect of applying $B_{-z-bias}$ on the magnetic tunneling barriers between the lattice sites, where the magnetic lattice is biased by $B_{x-biased}$ = 10 G. Simulation input: $\alpha_s=\alpha_h$ = 10 $\mu m$ $M_z$ = 2.80 kG, $\tau_{btm}$ = 2 $\mu m$ and $\tau_{p-wall}$  = 1 $\mu m$.}
\end{figure*}

As a result, each individual harmonic potential well is localized at its coordinate ($x_{min},y_{min},d_{min}$) and confined by magnetic barriers $\Delta\textbf{B}_{trap}$ where the barriers' heights can
be estimated to be

\begin{equation}
\Delta\textbf{B}_{trap}^{i}=|\textbf{B}_{max}^{i}|-|\textbf{B}_{min}^{i}|,\qquad i=x,y,z\label{eq:12}
\end{equation}

The magnetic fields around the local minima have gradients in which they are symmetrically distributed across the $x-y$ plane. These curvatures, along the $x$ and $y$ axes, are given by

\begin{eqnarray}
\frac{\partial^{2}\textbf{B}}{\partial x^{2}} &=& \Biggl(-\beta^{2}B_{ref}^{2}cos(\beta x) \nonumber \\
 & \times & \Biggl[\textbf{B}^{2}cos(\beta x)+B_{ref}cos(\beta x)sin^{2}(\beta x)\Biggl]\Biggr)/{\textbf{B}^{3}}  \nonumber \\ \label{eq:13}
\frac{\partial^{2}\textbf{B}}{\partial y^{2}} &=&  \Biggr(-\beta^{2}B_{ref}^{2}cos(\beta y)\nonumber \\
 & \times & \Biggl[\textbf{B}^{2}cos(\beta y)+B_{ref}cos(\beta x)sin^{2}(\beta y)\Biggl]\Biggr)/{\textbf{B}^{3}} \nonumber \\ \label{eq:14}
\end{eqnarray}

Due to the $x-y$ symmetry, we find that at the centers of the traps, the following holds true

\begin{equation}
\frac{\partial^{2}\textbf{B}}{\partial x^{2}}=\frac{\partial^{2}\textbf{B}}{\partial y^{2}}\label{eq:15}
\end{equation}

During the trapping process, before atoms reach the lowest magnetic field location at the bottom of traps, they undergo transverse oscillations.
The transverse oscillation frequency $\nu_{x,y}$ depends on the Zeeman sub-levels and the curvature of the magnetic field, and it can be expressed as

\begin{equation}
\nu_{k}=\frac{\beta}{2\pi}\sqrt{\mu_{B}g_{F}m_{F}\frac{\partial^{2}B_{k}}{\partial k^{2}}}\qquad\qquad\qquad k=x,y\label{eq:16}
\end{equation}

Where $g_{F}$ is the Land$\acute{e}$  g-factor, $\mu_{B}$ is Bohr magneton, $F$ is the atomic hyperfine state with the magnetic quantum number
$m_{F}$. The non-zero local minima values determine the depth of the harmonic potential wells, where the depth $\Lambda_{depth}$ of an individual
potential well can be expressed as folows

\begin{alignat}{1}
\Lambda_{depth}(\textbf{x}) & =\frac{\mu_{B}g_{F}m_{F}}{K_{B}}\Delta\textbf{B}_{trap}(\textbf{x})\nonumber \\
 & =\frac{\mu_{B}g_{F}m_{F}}{K_{B}}\Biggl(|B_{trap}^{max}(\textbf{x})|-|B_{trap}^{min}(\textbf{x})|\Biggl), \nonumber \\
\label{eq:17-1}
\end{alignat}

Where $\textbf{x} \equiv (x,y,z)$, $K_{B}$ is the Boltzmann constant, $B_{trap}^{max}$ and $B_{trap}^{min}$ are the maxima and minima values of the magnetic field barriers. Increasing the periodicity length, $\alpha_{s}$, between the holes increases the value of the non-zero local minima by several Gauss above the Majorana spin-flip critical minima. Our model shows suitable values for $\alpha_{s}$ that positively influence the cold atoms life time, where large values of $\alpha_s$ causes the magnetic minima to occur away from the surface, thereby reducing the interaction with the surface. The effect of changing the periodicity $\alpha_s$, can be used to locate suitable effective distances which keep the cold atoms away from the Casimir-Polder interacting limits when they are trapped near the surface of two-dimensional magnetic lattices, as shown in Figure \ref{Fig3}(a-b).

The gradient of the trapping magnetic field at each individual potential well is of particular importance when loading the ultracold atoms into the magnetic lattice. Lattice sites with a steeper gradient of their magnetic potentials maintain a suitable environment with increased chances for the destructive thermal Majorana spin-flip to exist. Consequently, the loss of ultracold atoms is more likely to occur due to the spin-flip process which causes an increase in their temperature. In this situation the in situ radio frequency (RF) evaporative cooling is effectively reducing the number of the re-thermalized (hot) atoms and hence enhance the phase space density of the compressed MOT to reach the point of BEC transition. On the other hand, steep gradients increase the chances of having single-atom per lattice site.  Moreover, the holes opening centers may reduce the effect of the surface Casimir-Polder interaction when using larger values of $\alpha_h$. Simulation results of various dimensions of  $\alpha_h$ and $\alpha_s$, for a magnetic lattice prepared at initial setup are shown in Figure (\ref{Fig4}). 


\begin{figure*}
\numberwithin{figure}{section}
\subfigure[]{
\label{Fig7a}}
\subfigure[]{
\label{Fig7b}}
\subfigure[]{
\label{Fig7c}}
\subfigure[]{
\label{Fig7d}}
\caption{\label{Fig7}(a) Scanning Electron Microscope (SEM) and (b) Atomic Force Microscope (AFM) images of the fabricated two-dimensional magnetic lattice. (c) The external magnetic field measured using the Magnetic Force Microscope (MFM) in which it agrees with the simulation result in (d) of the insitu biased two-dimensional magnetic lattice. Simulation and experimental inputs: $\alpha_s=\alpha_h$ = a0 $\mu m$ $M_z$ $\approx $ 2.80 kG, $\tau_{btm}$ = 1 $\mu m$ and $\tau_{p-wall}$  = 1 $\mu m$.}
\end{figure*}

 Our simulation results show that the unperturbed film area, of thickness $\tau_{wall}$, surrounding the $n\times n$ matrix affect the bottom of most of the potential minima at the center of the magnetic lattice. The effect of $\tau_{wall}$ (magnetic walls effect) on the magnetic lattice sites, introduces pronounce changes in the value of $B_{min}$ and $d_{min}$ at the edge and center lattice sites. To describe the effect of the magnetic walls on some of the characteristic lattice parameters, such as the magnetic tunneling barrier heights, $\Delta B$, and the effective distance, $d_{min}$, we simulate the trapping magnetic field above the surface of the magnetic lattice with different values of $\tau_{wall}$. Considering three values for $\tau_{wall}$ with respect to the bottom layer thickness $\tau_{btm}$, the terms positive wall, negative wall and surface-equal wall are introduced to denote the conditions $\tau_{p-wall}$, $\tau_{n-wall}$ and $\tau_{se-wall}$,  respectively. 

We summarize the two conditions positive wall, $\tau_{p-wall}$, and negative wall, $\tau_{n-wall}$, based on the number $n$ of the lattice sites, as follow. Applying the configuration $\tau_{p-wall}$ results in an unperturbed area with thickness that is larger than the thickness of the bottom layer of the magnetic lattice $\tau_{p-wall} \equiv\tau_{wall}>\tau_{btm}$, as shown in Figure \ref{Fig5}(a). For small values of $n$, the $\tau_{p-wall}$ acts as insitu magnetic bias field sources which surround the $n\times n$ magnetic lattice. Thus, $\tau_{p-wall}$ configuration with proper dimensions may replace the external magnetic bias fields sources created using current carrying wires in the BEC experiment. The effect of $\tau_{p-wall}$ appears as reduction in the amount of the tilting magnetic potential between sites, as shown in Figure \ref{Fig5}(b). Figure \ref{Fig5}(c) shows the locations of two different magnetic minima along $z$-axis at the center and the edge of the magnetic lattice where it is clear the $\tau_{p-wall}$ configuration reduces the spacial difference between sites.
For a large magnetic lattice, i.e., $n\rightarrow 100$, the $\tau_{p-wall}$ has no effect on the magnetic tunneling barriers and the spacial separations between sites at the center of the magnetic lattice. However, the edge sites are randomly distributed along the $z$-axis as shown in Figure \ref{Fig5}(d). It is also noticed that the $B_{min}$ of the center sites are close to zero and their $d_{min}$s are relatively small but still far from the surface.

When applying the condition $\tau_{n-wall}\equiv\tau_{wall}<\tau_{btm}$, the lattice sites are well spaced on the $z-x,z-y$ planes as shown in Figure \ref{Fig5}(e-f). Each $m$ pair of lattice sites around the center site have equal $B_{min}$, which led us to realize the existence of different magnetic bands surrounding the center site. We denote the effect by the magnetic band gap structure in a similar way to the energy band gap structure in semiconductor devices which can be used to simulate interesting problems in condensed matter physics using trapped ultracold atoms, e.g., ultracold fermions \cite{Ref_18}. Figure \ref{Fig5}(g), shows the locations of two different sites, i.e., $B_{min}$, along the $z$-axis at the center and the edge of the magnetic lattice. The pronounce spacial separation creates the magnetic band gap structure  in our proposed asymmetrical magnetic lattice.
The center of large magnetic lattice is also not affected in the case of $\tau_{n-wall}$ where center sites have $B_{min}$s close to zero while the edge sites maintain robust tilting magnetic potentials. Figure \ref{Fig5}(h) shows the simulation result of applying the condition $\tau_{n-wall}$ on a large magnetic lattice with $n$ = 100 sites.  

An asymmetrical magnetic lattice set initially at the reference magnetic field, $B_{ref}$, with no application of external bias fields, regardless of existence of the wall constrain, creates magnetic confinement with $B_{min}$ close to zero. This situation can be avoided by changing the periodicity between the lattice sites, i.e., changing the value of the separating period $\alpha_s$ between holes to be larger than the hole size  $\alpha_h$. In Table (\ref{Table3}) we summarize the simulation results of using different values for the hole size, i.e., $\alpha_h$ = 0.5 $\mu m$, 1 $\mu m$ and different values for the separating period, $\alpha_s$ = 0.5 $\mu m$, 1 $\mu m$, 1.5 $\mu m$, and 2 $\mu m$.

Another method of letting $B_{min} \ne $ 0 G, is to apply external magnetic bias fields which is commonly used in the area of magnetic micro-trapping of cold atoms. Figure \ref{Fig6} shows the effect of applying different magnetic bias fields where we simulate the case of $\alpha_h$ = $\alpha_s$ = 10 $\mu m$. For the tunneling process of the cold atoms to be effective periodicity with a relatively small value is desirable. A detailed analysis of achieving an optimal two-dimensional magnetic lattice, using this new method, will be reported elsewhere.

We fabricate the quantum device by milling the $n\times n$ hole structure using Focused Ion Beam (FIB) on the surface of magnetic thin film. Figure \ref{Fig7}(a-b) show 2D image and 3D image of the structure created by using Scanning Electron Microscope (SEM) and Atomic Force Microscope (AFM), respectively. The the external magnetic field of the asymmetrical magnetic lattice is measured using Magnetic Force Microscope (MFM), where the structure is biased insitu along the $x$-axis. Figure \ref{Fig7}(c-d) show the 3D image of the measured field and 3D plot of the simulation result, respectively. The experimental results will be reported elsewhere.


\section{\label{sec:5} Conclusion}

We have proposed a new method to create an asymmetric two-dimensional magnetic lattice suitable for trapping and confining ultracold atoms and quantum degenerate gases prepared in the low magnetic field seeking-state. The proposed remanent magnetic structure can be fabricated by milling an $m\times m$ array of blocks, where each block is an array of $n\times n$ square holes. Simulation results have shown that it is possible to optimize the two-dimensional magnetic lattice, i.e., to achieve symmetrical distribution of the lattice sites, via the application of external magnetic bias fields as well as by applying the suitable magnetic walls constrain. The many block feature will allow individual addressing for different quantum states which led us to coin the notion of multi-core quantum processors as will be reported elsewhere. 
We have also realized the possibility of creating discrete magnetic levels using the asymmetrical characteristic in the distributed magnetic field in which we have denoted the discretization effect by the magnetic band gap structure on atom chip similar to the energy band gap structure in semiconductor devices. This will allow, using the ultracold quantum degenerate gases, simulating condensed matter environment in which many interesting problems can be investigated such as exciton formation, Josephson effect and long-range entanglement. 

The spacing between the lattice sites is suitable for single atom trapping and detection where well spaced single atoms can be trapped at each site.
It is of a particular importance to highlight the simplicity of the method introduced in this article for achieving a two-dimensional magnetic lattice as it also suits the current state of the art technologies. We have shown that applying the external bias fields in the negative $z$-direction controls the tunneling process of cold atoms between sites as well as the in-site transitions. Thus, quantum information processing, such as multipartite entanglement, is also possible using such type of magnetic lattices.

\section{Acknowledge}  
We would like to thank James Wang from Centre for Atom Optics and Ultrafast Spectroscopy at Swinburne University in Melbourne, for his help in measuring the surface magnetic field using the Magnetic Force Microscope.

\bibliography{apssamp}

\end{document}